\documentclass[sigconf]{acmart}

\newcommand{\benchmark}{\textsc{EvalGPTFix}\xspace}
\newcommand{\gpt}{\text{ChatGPT}\xspace}

\usepackage{enumerate}
\usepackage{ragged2e}

\usepackage{stfloats}
\usepackage{amsmath,amsfonts}
\usepackage{algorithmic}
\usepackage{graphicx}
\usepackage{textcomp}
\usepackage{xcolor}
\usepackage{pifont}

\usepackage{url}
\usepackage{hyperref}
\hypersetup{
    colorlinks=true,
    linkcolor=blue,
    filecolor=blue,      
    urlcolor=blue,
    citecolor=blue,
}

\usepackage{multirow}
\usepackage[utf8]{inputenc}
\usepackage{algorithmic}
\usepackage{algorithm}
\usepackage{graphicx}
\usepackage{textcomp}
\usepackage{color,xcolor}
\usepackage{url}
\usepackage{graphicx}
\usepackage{subfigure}
\usepackage{stmaryrd}
\usepackage{verbatimbox}
\usepackage{enumerate}
\usepackage[shortlabels]{enumitem}
\usepackage{bm}%

\usepackage{blindtext}
\newenvironment{mylist}[1]%
{\begin{list}{}{\settowidth{\labelwidth}{\bf #1}%
			\setlength{\leftmargin}{\labelwidth}%
			\addtolength{\leftmargin}{\labelsep}%
			}}%
{\end{list}}

\usepackage{makecell}
\usepackage{booktabs}

\usepackage{multirow}
\usepackage{listings}

\usepackage{enumitem}
\usepackage{colortbl}
\definecolor{codegreen}{rgb}{0,0.6,0}
\definecolor{codegray}{rgb}{0.5,0.5,0.5}
\definecolor{codepurple}{rgb}{0.58,0,0.82}
\definecolor{backcolour}{rgb}{0.95,0.95,0.92}

\usepackage{listings}
\lstdefinestyle{mystyle}{
  language=Java,
  backgroundcolor=\color{backcolour},   
  commentstyle=\color{codegreen},
  keywordstyle=\color{magenta},
  numberstyle=\tiny\color{codegray},
  stringstyle=\color{codepurple},
  basicstyle=\ttfamily\scriptsize,
  breakatwhitespace=false,
  breaklines=true,
  captionpos=b,
  keepspaces=true,
  numbers=left,
  numbersep=5pt,
  showspaces=false,
  showstringspaces=false,
  showtabs=false,
  tabsize=2,
  frame = shadowbox,
}

\usepackage[most]{tcolorbox}

\newtcbox{\mylib}{enhanced,nobeforeafter,tcbox raise base,boxrule=0.4pt,top=0mm,bottom=0mm,
  right=0mm,left=4mm,arc=1pt,boxsep=2pt,before upper={\vphantom{dlg}},
  colframe=green!50!black,coltext=green!25!black,colback=green!10!white,
  overlay={\begin{tcbclipinterior}\fill[green!75!blue!50!white] (frame.south west)
    rectangle node[text=white,font=\sffamily\bfseries\tiny,rotate=90] {TYP} ([xshift=4mm]frame.north west);\end{tcbclipinterior}}}

\newtcolorbox{mybox}[2][]{colback=gray!15, colframe=black,fonttitle=\bfseries,  enhanced, attach boxed title to top center={yshift=-2mm}, breakable, left=2mm,right=2mm, bottom=1mm, title={$\blacktriangleright$#2$\blacktriangleleft$}, #1}

\newcommand{\myfinding}[2]{
\begin{center}
\begin{tcolorbox}[colback=gray!15, colframe=black, boxsep= -0.15cm, middle=-0.15cm, breakable]
\textbf{Answer to RQ{#1}:}
{#2}
\end{tcolorbox}
\end{center}
}

\usepackage{xspace}
\newcommand{\ie}{\textit{i.e.,}\xspace}
\newcommand{\eg}{\textit{e.g.,}\xspace}

\newcommand{\etal}{\textit{et al.}\xspace}

\usepackage[normalem]{ulem}
\newcommand{\revise}[1]{{\color{black}{#1}}}
\newcommand{\delete}[1]{}

\AtBeginDocument{%
  \providecommand\BibTeX{{%
    \normalfont B\kern-0.5em{\scshape i\kern-0.25em b}\kern-0.8em\TeX}}}

\acmConference[Conference acronym 'XX]{Make sure to enter the correct
  conference title from your rights confirmation emai}{June 03--05,
  2023}{Woodstock, NY}
\acmPrice{15.00}
\acmISBN{978-1-4503-XXXX-X/23/06}

\begin{document}

\title{A Critical Review of Large Language Model on Software Engineering: An Example from ChatGPT and Automated Program Repair}

\author{Quanjun Zhang}    \email{quanjun.zhang@smail.nju.edu.cn}
\affiliation{
  \institution{State Key Laboratory for Novel Software Technology, Nanjing University}
  \city{Nanjing}
  \state{Jiangsu}
  \country{China}
  \postcode{210093}
}

\author{Tongke Zhang}    \email{201250032@smail.nju.edu.cn}
\affiliation{
  \institution{State Key Laboratory for Novel Software Technology, Nanjing University}
  \city{Nanjing}
  \state{Jiangsu}
  \country{China}
  \postcode{210093}
}

\author{Juan Zhai} \email{juanzhai@umass.edu}
\affiliation{
  \institution{Manning College of Information \& Computer Sciences, University of Massachusetts}
  \city{Amherst}
  \state{MA}
  \country{United States}
  \postcode{01003}
}

\author{Chunrong Fang} \email{fangchunrong@nju.edu.cn}
\authornote{\textbf{Chunrong Fang is the corresponding author.}}
\affiliation{
  \institution{State Key Laboratory for Novel Software Technology, Nanjing University}
  \city{Nanjing}
  \state{Jiangsu}
  \country{China}
  \postcode{210093}
}

\author{Bowen Yu}    \email{201250070@smail.nju.edu.cn}
\affiliation{
  \institution{State Key Laboratory for Novel Software Technology, Nanjing University}
  \city{Nanjing}
  \state{Jiangsu}
  \country{China}
  \postcode{210093}
}

\author{Weisong Sun} \email{weisongsun@smail.nju.edu.cn}
\affiliation{
  \institution{State Key Laboratory for Novel Software Technology, Nanjing University}
  \city{Nanjing}
  \state{Jiangsu}
  \country{China}
  \postcode{210093}
}

\author{Zhenyu Chen} 
\email{zychen@nju.edu.cn}
\affiliation{
  \institution{State Key Laboratory for Novel Software Technology, Nanjing University}
  \city{Nanjing}
  \state{Jiangsu}
  \country{China}
  \postcode{210093}
}

\begin{abstract}

Large Language Models (LLMs) have been gaining increasing attention and demonstrated promising performance across a variety of Software Engineering (SE) tasks, such as Automated Program Repair (APR), code summarization, and code completion.
For example, ChatGPT, the latest black-box LLM, has been investigated by numerous recent research studies and has shown impressive performance in various tasks.
However, there exists a potential risk of data leakage since these LLMs are usually close-sourced with unknown specific training details, \eg pre-training datasets.

In this paper, we seek to review the bug-fixing capabilities of {\gpt} on a clean APR benchmark with different research objectives.
We first introduce {\benchmark}, a new benchmark with buggy and the corresponding fixed programs from competitive programming problems starting from 2023, after the training cutoff point of {\gpt}.
The results on {\benchmark} show that {\gpt} is able to fix 109 out of 151 buggy programs using the basic prompt within 35 independent rounds, outperforming state-of-the-art LLMs CodeT5 and PLBART by 27.5\% and 62.4\% prediction accuracy.
We also investigate the impact of three types of prompts, \ie problem description, error feedback, and bug localization, leading to additional 34 fixed bugs.
Besides, we provide additional discussion from the interactive nature of {\gpt} to illustrate the capacity of a dialog-based repair workflow with 9 additional fixed bugs.
Overall, our experiments demonstrate that {\gpt} is able to fix a total of 143 bugs in {\benchmark}, indicating the potential of {\gpt} in repairing real-world buggy programs. 
Inspired by the findings, we further pinpoint various challenges and opportunities for advanced SE study equipped with such LLMs (\eg~{\gpt}) in the near future.
More importantly, our work calls for more research on the reevaluation of the achievements obtained by existing black-box LLMs across various SE tasks, not limited to \gpt{} on APR.
\end{abstract}



\maketitle

\section{Introduction}
\label{sec:intro}

The scale of modern software systems has been continuously expanding in recent years, leading to a significant surge in the number of bugs within these systems~\cite{monperrus2018automatic,gazzola2019automatic}. 
Manual debugging is one of the critical software development activities, which usually requires a substantial investment of time and human resources to keep the software well-maintained~\cite{benton2020effectiveness,britton2013reversible}.
In order to reduce the costs of repairing such detected bugs, Automated Program Repair (APR) is proposed~\cite{le2012genprog}, aiming at generating correct patches automatically to minimize human involvement in the manual debugging process.

In the literature, existing APR techniques can be categorized into traditional and learning-based ones~\cite{monperrus2020living,gao2022program}. 
Among traditional APR techniques, template-based APR, which mainly relies on pre-defined repair templates to transform buggy code into the correct one, has been considered as state-of-the-art~\cite{xia2022practical,xia2022less}.
However, it is challenging to fix unseen bugs that fall outside the scope of pre-defined templates.
On the other hand, learning-based APR is able to learn the bug-fixing patterns automatically from a large code repository on top of the advance of deep learning~\cite{chen2022neural,chi2022seqtrans}. 
Learning-based APR usually leverages Neural Machine Translation (NMT) model to translate a code sequence from a source language (i.e., buggy code snippets) into a target language (i.e., correct code snippets)~\cite{tufano2019learning}.
Despite addressing the limitations of template-based APR and demonstrating promising results, the performance of learning-based APR relies on the quality and quantity of the training data~\cite{xia2022less}.

More recently, Large Language Models (LLMs) are gaining increasing attention due to their powerful programming language processing capabilities in various Software Engineering (SE) tasks~\cite{xia2022practical,fu2022vulrepair,mastropaolo2021t5learning}.
These LLMs are usually trained with a pre-training-and-fine-tuning mechanism~\cite{wang2021codet5,feng2020codebert}, \ie pre-trained by self-supervised training on a large-scale unlabeled corpus to derive generic knowledge, and fine-tuned by supervised training on a limited labeled corpus to benefit a specific downstream task.
Among existing LLMs, ChatGPT~\cite{chatgpt} is widely regarded as one of the most popular language models today and is being studied by researchers from numerous domains, such as code summarization~\cite{sun2023automatic}, code generation~\cite{liu2023improving} and test generation~\cite{yuan2023no}.
In particular, ChatGPT is a prompt-based LLM equipped with Reinforcement Learning from Human Feedback that can interact with users through human-like dialogues. 
In the domain of APR, researchers have attempted to directly utilize {\gpt} in generating correct patches and have yielded promising results~\cite{sobania2023analysis,xia2023keep}.
For example, Sobania~\etal~\cite{sobania2023analysis} evaluate the bug-fixing capabilities of ChatGPT and find \gpt{} is able to fix 31 out of 40 bugs on QuixBugs benchmark. 
Xia~\etal~\cite{xia2023keep} employ ChatGPT in a conversational manner to fix 114 and 48 bugs on the Defects4J-v1.2 and Defects4J-v2.0 benchmark, and all the 40 bugs from QuixBugs benchmark, significantly outperforming state-of-the-art APR techniques.

\begin{figure*}[htbp]
    \centering
    \includegraphics[width=0.9\linewidth]{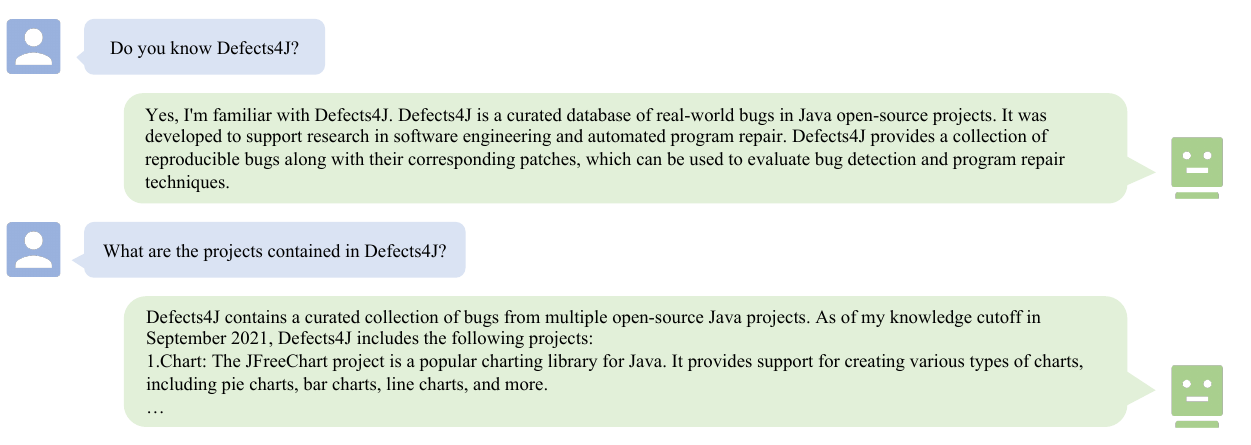}
    \caption{Dialogue with ChatGPT about its knowledge of Defects4J Benchmark}
    \label{fig:aware_of_d4j}
\end{figure*}

In spite of the remarkable performance, there are some concerns with the well-known dataset used to evaluate \gpt{} for APR.
\gpt{} is trained on vast amounts of data from the internet, which may contain data in the commonly-chosen datasets for APR (\eg Defects4J~\cite{just2014defects4j} and QuixBugs~\cite{lin2017quixbugs}).
It is difficult to ensure whether or not the evaluation dataset has not been seen by ChatGPT during training.
For example, when we ask ChatGPT whether it is aware of Defects4J, as shown in Figure~\ref{fig:aware_of_d4j}, {\gpt} gives an affirmative answer and can further list the projects present in Defects4J.
If ChatGPT has previous knowledge of the dataset, employing it to fix bugs from the dataset might not well reflect its fixing ability since it may already be aware of the bug-fixing patches.
The concern exists in other LLMs and code-related tasks as well.
A similar example shows\footnote{\url{https://github.com/hitz-zentroa/lm-contamination}} that GPT-4 is able to solve all 10 code competition problems until 2021, which is the training cutoff of the model, while none is solved correctly for 10 problems after that date instead.
We also find that {\gpt} can directly provide complete descriptions and the corresponding solution by simply providing it with the number of a programming problem in LeetCode \revise{(presented in our online repository~\cite{myurl})}.
Similarly, Karmakar~\etal~\cite{karmakar2022codex} highlight the memorization issue of Codex, showing its ability to generate accurate code outputs only with the first sentence of the problem description as a prompt, even in the absence of clear task objectives.
Considering the fact there exist a quite number of black-box LLMs for which no architecture or training data information has been released.
The data leaking on such LLMs is a significant concern when it comes to evaluating their performance in some code-related tasks in the SE community, such as APR~\cite{xia2022practical} and assertion generation\cite{nashid2023retrieval}.

\textbf{This paper.}
In our work, we attempt to raise the important concern about data leakage, which has been an overlooked issue in the SE community, when such black-box LLMs are applied to some code-related tasks.
We select {\gpt} and APR as representative examples of LLMs and SE tasks, respectively.
In particular, we construct a new dataset {\benchmark} with buggy and correct code nippets from a competitive programming website Atcoder.
We crawl users' submissions for competitions during 2023.
As ChatGPT states its knowledge cutoff is in September 2021, thus we confirm it does not have access to samples in {\benchmark}.
We then design the following three research questions to evaluate ChatGPT's bug-fixing ability in our experiments:

\begin{mylist}{\textbf{(RQ1)}}     

    \item[\textbf{(RQ1)}] \textbf{The effectiveness of {\gpt} on {\benchmark}.}
  
    \textbf{\underline{Results:} }
    In this RQ, we aim to investigate how ChatGPT behaves in repairing bugs when we present it with the buggy code.
    Although ChatGPT could learn more from detailed prompts, we only give a prompt containing the bug and an instruction that simply asks to fix the bug. This is similar to the input for common APR methods, so we can compare the repair effectiveness of ChatGPT with other tools.
    We require ChatGPT to try to fix 151 bugs chosen from two latest AtCoder competitions in {\benchmark}, and count the number of bugs that ChatGPT is able to fix to explore to what extent can ChatGPT generate correct patches.

    \item[\textbf{(RQ2)}] \textbf{What is the effect of different prompts on the repair performance of ChatGPT?}
  
    \textbf{\underline{Results:} }
    ChatGPT is a prompt-based language model, whose response is largely dependent on how it is prompted. Through modifying prompts, we can give more information relating to the bugs. After gaining such information, ChatGPT has the potential to fix more bugs. To find out what prompt helps in bug fixing, we design three advanced prompts which separately include the programming problems that the programs target at, the exact lines where the bugs are in, and what type of error the bugs have. We apply these prompts to the bugs not fixed in RQ1 and experiment on how many more bugs can be fixed by each prompt.
    
    \item[\textbf{(RQ3)}] \textbf{Can dialogues help ChatGPT in improving repair performance?}
  
    \textbf{\underline{Results:} }
    ChatGPT is created to engage in dialogues with users, and while answering a query, it has a memory of previous conversations, from which it can adjust its responses. 
    For the bugs that ChatGPT fails to fix when only a single round of dialogue is performed, we continue the dialogues by informing ChatGPT of what is wrong with the patches it has generated. 
    The intuition is that ChatGPT can learn from the implausible fixes and gradually turn its attention to the correct patch with the accumulative dialogues.
\end{mylist}

Overall, our work confirms that ChatGPT has an excellent performance in fixing bugs from the dataset {\benchmark}. 
In the case of providing the basic prompt, ChatGPT fixes 109 bugs out of the 151 bugs. 
When programming problem descriptions, error messages and bug locations are added to the prompt, respectively 18, 25 and 10 more bugs are fixed. 
Moreover, by conducting dialogues, ChatGPT fixes 9 bugs that are neither fixed with the basic prompt nor with the prompt including error information. 
The results show that the repair ability of ChatGPT can benefit from prompts that are more in detail and deeper dialogues.

\textbf{Novelty \& Contributions.}
To sum up, the main contributions of this paper are as follows:

\begin{itemize}
\item
\textit{\textbf{Overlooked Issue.}}
We reveal an important concern when evaluating the recent \gpt{} in repairing software bugs with commonly-adopted benchmarks, i.e., \textit{the data leakage issue}.
More importantly, the issue potentially exists in a broader range of other code-related tasks and black-box LLMs, and has been consistently overlooked by the SE community.
Thus, the issue may lead to a significant bias in previous research works that employ black-box LLMs (such as {\gpt} and Codex) without assessing any training details, such as pre-training datasets and model architectures.

\item
\textit{\textbf{Clean Benchmark.}}
We \revise{construct} a new APR benchmark \benchmark{} from a competitive programming website Atcoder.
\revise{\benchmark{} contains 151 pairs of bugs and fixes in Java, which come from failing and accepted programming submissions in 2023 to ensure that \gpt{} has not seen the specific code snippets presented in this dataset.}

\item
\textit{\textbf{Extensive Study.}}
We conduct an in-depth empirical study of how \gpt{} are applied to automated program repair.
Specifically, our study is three-fold: 
(1) a systematic comparison between \gpt{} and state-of-the-art LLMs, indicating that {\gpt} can outperform CodeT5 and PLBART;
(2) an extensive evaluation to analyze the impact of different prompts;
(3) an additional discussion about the impact of dialogue-based repair workflow for \gpt{}.

\item
\textit{\textbf{Challenge and Opportunity}}
We discuss current pressing challenges and  forward-looking directions on applying \gpt{} and more advanced black-box LLMs for future program repair and other SE studies.

\end{itemize}  

\textbf{Open Science.}
To support the open science community, we release the studied dataset, scripts (i.e., data processing, model training, and model evaluation), and related models in our experiment for replication and future research~\cite{myurl}.

\section{Background}
\label{sec:bg&mv}
\subsection{Automated Program Repair}
Automated Program Repair (APR) is raised to generate candidate patches automatically for the buggy code snippets, so as to reduce the time and cost of manual debugging~\cite{xiong2017precise,zhu2023tare}.
There are mainly two types of APR techniques, \ie traditional and learning-based ones.

Traditional APR can be classified into three categories: heuristic-based~\cite{le2012genprog, martinez2016astor,yuan2018arja}, constraint-based~\cite{martinez2018ultra,durieux2016dynamoth, mechtaev2016angelix}, and template-based~\cite{koyuncu2020fixminer,Liu2019Avatar,liu2019tbar}. 
Among them, template-based APR has shown promising performance in the bug-fixing task. 
Template-based APR utilizes pre-defined fix templates, which are patterns of code changes commonly applied in debugging activities, to generate possible patches for specific bugs. 
Despite of its significant ability in program repair, template-based APR has limitations in both fix templates and donor code.
It cannot fix bugs requiring action beyond the fix templates, and with a proper fix template, some bugs still cannot be fixed because of a limited source of donor code~\cite{liu2019tbar}. 

The problems above can be solved by learning-based APR, which is a rising field that APR researchers are focusing on. 
APR problems are regarded as Neural Machine Translation (NMT) tasks that transform a piece of buggy code into the correct one. 
Learning-based APR leverages DL techniques to gain more insight into program repair behaviors from large code corpora. 
However, the effectiveness of learning-based APR can be easily affected by the quality of training data, which may contain code changes irrelevant to bug fixing and thus limit the performance of APR.

\subsection{Large Language Model}
Large Language Models (LLMs) are language models consisting of billions of parameters trained from a significant amount of data and have impressive performance in language processing tasks, including both natural languages and programming languages~\cite{wang2021codet5,feng2020codebert,guo2020graphcodebert,guo2022unixcoder,karmakar2022codex}. 
LLMs are typically based on Transformer architecture, where an encoder takes a variable-length input and turns it into a fixed-length vector, while a decoder transforms the encoded representation into a sequence of output. 
Based on the employed components, LLMs can be categorized into encoder-only, decoder-only, and encoder-decoder ones. 
Encoder-only models (\eg BERT\cite{devlin2018bert}) learn data representation through training objectives like Masked Language Modeling (MLM).
Decoder-only models (\eg GPT\cite{brown2020gpt}) are trained to generate predictions for the next token given all the previous tokens. 
Encoder-decoder models (\eg CodeT5 \cite{wang2021codet5}) combine the encoder and the decoder, and are trained with the objective of recovering the corrupted input.

\section{Study Design}
\label{sec:sd}

\subsection{Research Questions}

In this work, we explore the following research questions.

\begin{description}
  \item [RQ1:] What is the performance of {\gpt} in repairing buggy programs from {\benchmark}?

  \item [RQ2:] How do the different prompts with additional program information affect the performance of {\gpt}?
  
  \item [RQ3:] How does the interaction with dynamic execution feedback affect the performance of  {\gpt}?
\end{description}

\subsection{{\benchmark} Construction}
\label{sec:dataset}

There is evidence showing that ChatGPT already has knowledge about popular datasets (\eg Defects4J and Quixbugs) widely used by current APR techniques. 
As a result, it seems not rigorous to research on ChatGPT's program repair ability based on these datasets as ChatGPT may find a correct patch for the given bug from its training data instead of fixing the bug by itself. 
To solve this problem, we construct a new dataset whose data is invisible to ChatGPT. We gain our data from AtCoder, a platform for programming contests. We extract the programming problems from contests in 2023 and get users' submissions of these problems as the source data. As ChatGPT is trained on data before September 2021, it has limited knowledge about our dataset, so applying ChatGPT to fix bugs from \benchmark can better reflect ChatGPT's performance in the APR task.

\textbf{\ding{172} Raw Data Collection.}
We first crawl all the Java submissions of AtCoder programming contests starting from 2023.
We focus on Java languages as it is the most targeted language in the APR community.
The online judge results of the submissions can be divided into six types: 1) Accepted (AC); 2) Wrong Answer (WA); 3) Time Limit Exceeded (TLE); 4) Compilation Error (CE); 5) Runtime Error (RE); and 6) Memory Limit Exceeded (MLE). 

\textbf{\ding{173}~Bug-Fixing Pairs Construction.}
For all the submissions of a user on a single problem, we take the unaccepted submissions as the buggy program and the accepted submission as a corresponding correct program. 
Then we calculate the token difference between every pair of the buggy and correct programs, and only keep those with a difference of less than 6 tokens following existing the study~\cite{haque2022fixeval}.
\revise{This difference is calculated by tokenizing each program into a sequence of tokens and counting the number of token-level differences, including replacements, deletions, and insertions, between the two programs.}
This setting is based on the \textit{competent programmer hypothesis}, \ie seasoned programmers have the competence to produce programs that are nearly error-free, and that the majority of bugs can be rectified through minor modifications.

\textbf{\ding{174}~Test Case Mining.}
For each problem, the problem description in the HTML website includes a handful of illustrative input-output pairs that serve as examples.
However, these test cases are not sufficient to validate the correctness of generated patches due to the overfitting problem in APR~\cite{zhang2023boosting}.
We further download all possible public test cases of all the problems in our dataset from the dedicated database\footnote{\url{https://www.dropbox.com/sh/arnpe0ef5wds8cv/AAAk_SECQ2Nc6SVGii3rHX6Fa?dl=0}} that posts all test cases of AtCoder problems.
These test cases are manually created by domain experts, such as the programming problem setters, and serve as the test oracle in the backend of the website to assess the functional correctness of programs submitted by users.

\textbf{\ding{175}~Static-based Filtering.}
\revise{Furthermore, we remove repeating submissions as well as submissions with more than 500 code tokens, considering the limitation of repair models' ability to handle long code snippets (\eg Tufano~\etal~\cite{tufano2019empirical} limit the maximum length of the buggy code to 100 tokens and CIRCLE~\cite{yuan2022circle} truncates the first 512 tokens of the buggy code}).
We also delete the comments in the code, as comments can provide information about the function of the buggy code, and can affect the judgment of APR tools' capability in fixing code without other hints. 
We also observe that some programs have a compilation error only because the class name is not written as \texttt{"Main"}. 
This is irrelevant to the logic of the code itself, so such data is deleted from our dataset.

\textbf{\ding{176}~Dynamic-based Filtering.}
We execute all remaining submissions against every test case associated with the respective problem.
The time limit and a memory limit of running a test case are respectively set as 10 seconds and 1MB \revise{following previous work~\cite{prenner2022can}}. 
In a pair $(s1, s2)$ where $s1$ represents the buggy code and $s2$ represents the fixed one, if any of the following three conditions happen, the pair will be removed:
(1) $s1$ passes all the test cases of its corresponding problem;
(2) The bug type of $s1$ does not match the one given on AtCoder website (\eg $s1$ is found to produce a "Wrong Answer" error by running test cases but it gets a label of "Compilation Error" on AtCoder);
and (3) $s2$ fails any of the test cases of its corresponding problem.
Besides, due to differences in our local device and AtCoder platform environments, we exclude bugs with a type of MLE, resulting in four types of bugs in \benchmark (\ie WA, TLE, CE, and RE). 

\textbf{\ding{176}~Benchmark Statistics.}
After all pre-processing phrases, we get 151 pairs of bug-fixing of Java programs for 15 programming problems from the two latest programming contests when we conduct the work, \ie Beginner Contest 297 and 298.

\subsection{ChatGPT Setup}
We conduct our experiment based on the API of ChatGPT with the model \texttt{gpt-3.5-turbo} released by OpenAI. 
Considering that ChatGPT will generate different responses when it is queried by the same input several times, with every prompt we send the request to ChatGPT repeatedly to improve the possibility of generating more correct patches.

\subsection{Evaluation Metrics}
\label{sec:metric}
We evaluate whether the patch generated by ChatGPT is correct by running it against the test suite. 
All the test cases are downloaded from the website of AtCoder, and they are used to judge users' submissions during coding contests. 
Averagely, every programming problem has 38 test cases, which is enough to tell whether the program can correctly solve the problem.
We run every candidate fix on the test cases of the problem, and if it passes all the test cases, it is regarded as a correct fix. 

For every bug with a prompt, we ask ChatGPT for a fix continuously in a loop, and if in any of the rounds, ChatGPT can generate a correct patch for a bug, we consider ChatGPT to be able to fix the bug successfully, and the loop will be exited.

\subsection{Compared Techniques}

We consider the following two state-of-the-art LLMs as the baseline techniques.

\textbf{$\bullet$ CodeT5.}
Wang~\etal~\cite{wang2021codet5} introduce a pre-trained language model (\ie CodeT5) on top of the T5 architecture by incorporating the token type information.
CodeT5 considers both unimodal (code only) and bimodal (code-text pairs) data for four pre-training tasks, \ie masked span prediction, masked identifier prediction, identifier tagging, and bimodal dual generation.

\textbf{$\bullet$ PLBART.}
Ahmad~\etal~\cite{ahmad2021unified} introduce a pre-trained encoder-decoder model (\ie PLBART) on top of the BERT architecture to perform both program and language understanding and generation tasks.
PLBART considers three denoising auto-encoding strategies to reconstruct an input text that is corrupted by a noise function in pre-training, \ie token masking, token deletion, and token infilling.

\section{Results and Analysis}
\label{sec:re&an}

\subsection{RQ1: Effectiveness of {\gpt}}
\label{sec:rq1}

\textbf{\emph{Design.}}
In this RQ, we explore the repair ability of ChatGPT by presenting it with buggy programs and asking it to repair them. 
We only give the buggy code without any other information about the bugs to find out to what extent can ChatGPT localize and repair bugs if no extra prompts are provided. 
The basic prompt designed for RQ1 is presented as follows, where [CODE] represents the buggy program to be fixed.

\begin{mybox}{Basic Prompt}
There's a bug in the program below. Try to fix it and return the complete fix for the code in the form of the markdown code block.

\ding{45}~[CODE]
\end{mybox}

ChatGPT generates different responses even if it is prompted by the same sentences, so if it is not able to return the correct patch for a bug, there is a possibility that it will fix the bug when queried again. 
Therefore, asking ChatGPT to fix a bug only once cannot reflect its actual repair capability well. 
To solve this problem, for every bug, we repetitively send the same request to ChatGPT. 
If a patch for a bug can pass all the test cases, we stop asking ChatGPT to fix the bug again. 
In every round of query, we check whether any new bugs are fixed compared to the last round, and if no more bugs are fixed for three consecutive rounds, the process is stopped.

\begin{figure}[htbp]
    \centering
    \includegraphics[width=0.9\linewidth]{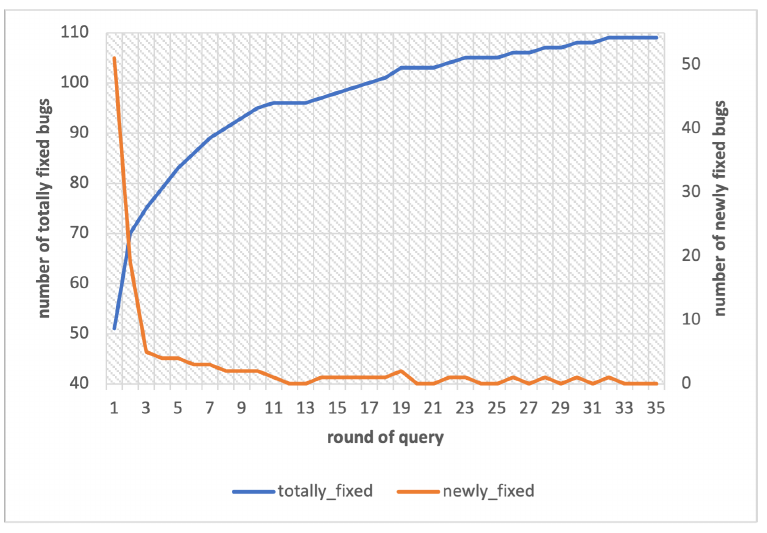}
    \caption{The number of totally fixed and newly fixed bugs in every round of query to ChatGPT}
    \label{fig:rq1}
\end{figure}

\begin{figure}[htbp]
    \centering
    \includegraphics[width=0.8\linewidth]{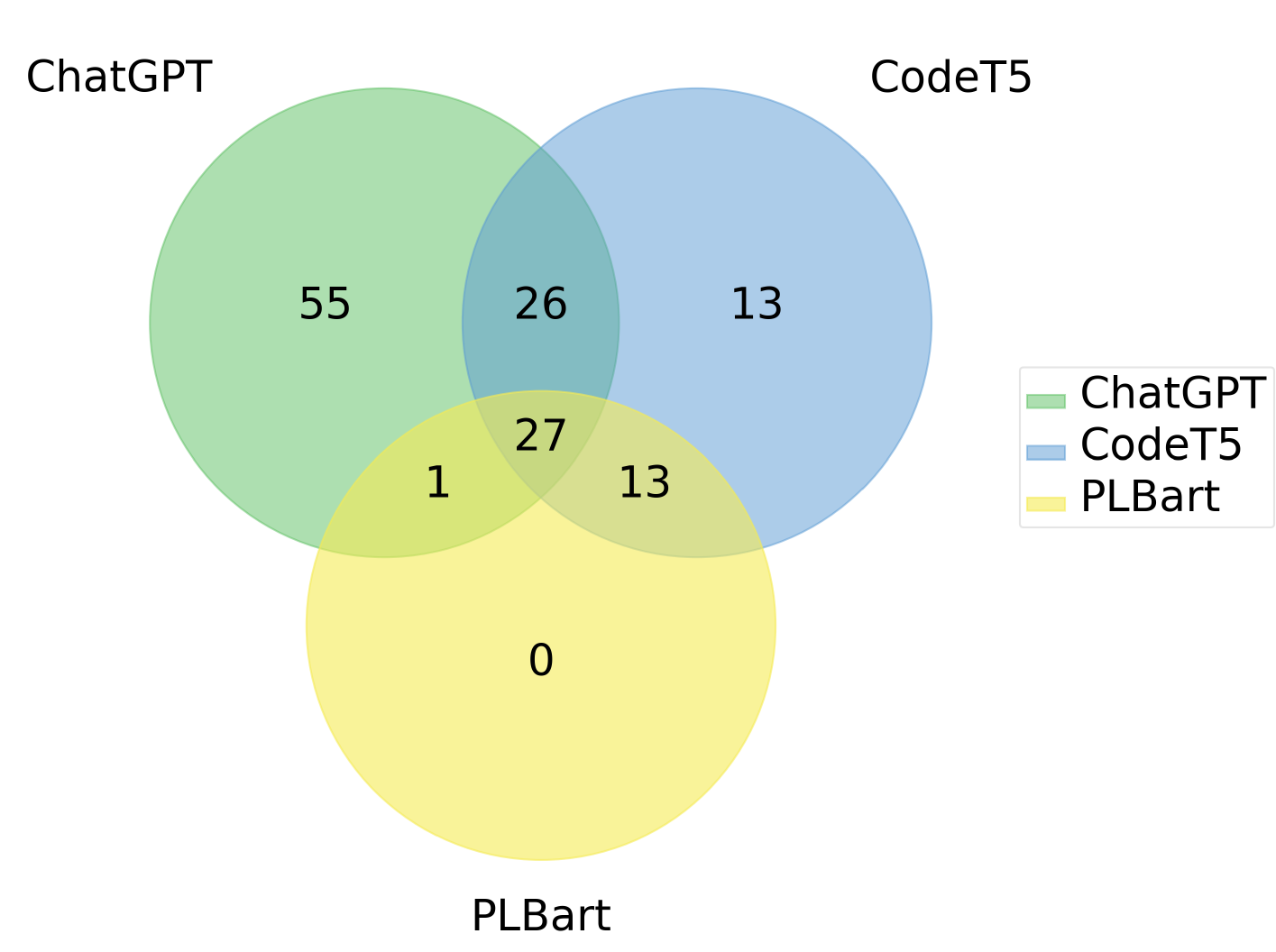}
    \caption{The overlap of bugs fixed by three models}
    \label{fig:models_venn}
\end{figure}

\textbf{\emph{Results.}}
Figure \ref{fig:rq1} shows the number of bugs that {\gpt} is able to fix in each round of requests.
We finally query ChatGPT for 35 independent rounds, in which the 33-35 rounds do not see any newly-fixed bugs. 
Among the 151 bugs from the two latest programming contests at AtCoder, 109 are successfully fixed. 
Generally, there is a decline in additional fixes with the increase of round number. 
In the first three rounds, respectively 51, 19, and 5 additional correct patches are generated, which is a sharp decrease.
After that, the number begins to drop slowly, and vacillates between 0 to 2 since the 11th round.
The decreasing trend only stopped after 35 rounds, indicating the randomness nature of {\gpt}.
Therefore, we recommend future work to explore the issue of randomness in {\gpt}, which has been ignored in most previous studies, \eg Sobania~\etal~\cite{sobania2023analysis} only repeats four times.

We then invesgiate the types of bugs fixed by {\gpt}.
Among the 151 bugs in {\benchmark}, there exist 23, 4, 22, and 102 bugs of the four types CE, TLE, RE, and WA.
{\gpt} is able to fix 22, 4, 11, and 72 of them, with the fixing percentage of 96\%, 100\%, 50\%, and 71\%.
We find ChatGPT shows an impressive performance in localizing and fixing bugs with Compilation Error(CE), which is usually caused by syntax errors. 
Such errors are easy to identify as long as ChatGPT has knowledge of Java syntax, so it can fix the bug without considering the operating logic of the program, which can be much more intricate.
For example, in the following bug \revise{shown in Listing~\ref{listing:rq1}}, the array operator "[]" is used on a string, which is not allowed in Java. ChatGPT recognizes the error and replaces \texttt{"S[i]"} with \texttt{"S.charAt(i)"}, and gives a correct patch.

\begin{lstlisting}[caption={An example of bug with syntax error fixed by basic prompt},label={listing:rq1},float]
    ...
    String S = sc.next();
    boolean f1 = false;
    boolean f2 = true;
    for(int i=0;i<N;i++){
-       if(S[i]=='o'){
+       if(S.charAt(i)=='o'){
          f1 = true;
        }
-       if(S[i]=='x'){
+       if(S.charAt(i)=='x'){
          f2 = false;
        }
    }
    ...
\end{lstlisting}

Apart from simple syntax errors, ChatGPT can also identify and repair some logic errors in the programs. 
Listing~\ref{listing:le} presents an example with a logic error that can be successfully fixed by {\gpt}.
In the buggy program, the conditional expression \texttt{"i<t.length"} in the while statement is problematic, as in the next statement it can trigger \textit{ArrayIndexOutOfBoundsException}, which is a kind of Runtime Error. ChatGPT changes the buggy expression into \texttt{"i<t.length-1"} to prevent the exception, so as to solve the problem in the code snippet.

\begin{lstlisting}[caption={An example of bug with logic error fixed by basic prompt},label={listing:le},float]
    ...
    int[] t = new int[n];
    ...
-   while(i<t.length)
+   while(i<t.length-1)
    {
      if(t[i+1]-t[i]<=d)
      {
    ...
\end{lstlisting}

We further compare the repair performance of ChatGPT with the other two state-of-the-art pre-trained models, \ie CodeT5 and PLBART. 
The models are fed with the buggy code and then generate possible patches which are later run on the test cases. 
We fine-tune the selected models with the FixEval dataset~\cite{haque2022fixeval}, which contains 156k buggy and correct code submissions to competitive programming problems before 2021. 
The beam size is set as 50 for both models, which means 50 patches with highest possibility are produced for each bug. 
We find that CodeT5 fixes 79 bugs and PLBART only fixes 41 bug, 27.5\% and 62.4\% less than what is achieved by ChatGPT. 
Figure \ref{fig:models_venn} presents the overlap of bugs fixed by the three models, showing that ChatGPT significantly fixes more bugs (\ie 55 unique bugs) that the other two models are not able to fix (\ie 13 unique bugs for CodeT5 and none for PLBART). 
The results indicate that ChatGPT notably outperforms existing LLMs with regard to repairing programming problems.

\myfinding{1}{
The performance of {\gpt} in {\benchmark} shows that:
(1) there exists a significant randomness issue in {\gpt}, \eg 35 independent rounds are required to achieve stable results;
(2) {\gpt} is effective in fixing different types of bugs, \eg \revise{
96\%, 100\%, 50\% and 71\% of CE, TLE, RE and WA bugs are correctly fixed;}
(3) {\gpt} is able to fix 109 bugs in {\benchmark} with a recall of 72.19\%, 30 and 68 more than CodeT5 and PLBART.
}

\subsection{RQ2: The Impact of Prompt}
\label{rq2}
\textbf{\emph{Design.}}
We further add more details about the bug to the prompt given to ChatGPT, expecting that ChatGPT will gain more useful information from the prompts so that it can fix more bugs.
We ask ChatGPT to fix the bugs that are not fixed when only the basic prompt is offered, as described in RQ1.
We consider three types of additional bug information, including problem descriptions, error information, and bug locations. 
For the three types of bug information, we respectively design three kinds of prompts based on the original prompt, discussed as follows.

\textbf{$\bullet$~Problem descriptions} indicate what the programming problem aims to solve. 
The prompt with a problem is shown below, where [CODE] represents the buggy program and [PROBLEM] represents the coding problem that the code is submitted to. 
All the problem descriptions are obtained from the website of AtCoder, and consist of the background of the problem, the input to the program, and what the output is supposed to be like.

\begin{mybox}{Problem Description Prompt}
\ding{46}~There's a bug in the program below. Try to fix it and return the complete fix for the code in the form of the markdown code block.

\ding{45}~[CODE]

The program is to solve this problem:

\ding{45}~[PROBLEM]
\end{mybox}

\textbf{$\bullet$~Error information} informs ChatGPT of the type of error, the input that triggers the error, the expected correct output, and the actual wrong output of the buggy program.
For the four types of bugs (i.e. CE, TLE, RE, and WA), the prompts are slightly different from each other, displayed as follows.

\ding{182}~Compilation Error (CE) is not triggered by any input as it happens while compiling, so in the case of a CE bug, we only tell ChatGPT that there is a compilation error in the code. 

\begin{mybox}{Compilation Error Prompt}
There's a bug in the program below. Try to fix it and return the complete fix for the code in the form of the markdown code block.

\ding{45}~[CODE]

There's a Compilation Error in the code.
\end{mybox}

\ding{183}~For a Time Limit Exceeded (TLE) or Runtime Error (RE), the input that causes the error as well as the expected output are added to the prompt.
The prompt is described as follows, where [INPUT] represents the input that causes the test case to fail and [EXPECT] represents the output that should be printed by a correct program.

\begin{mybox}{Exceeded/ Runtime Error Prompt}
There's a bug in the program below. Try to fix it and return the complete fix for the code in the form of the markdown code block.

\ding{45}~[CODE]

The following input triggers a Time Limit Exceeded/ Runtime Error:

\ding{45}~[INPUT]

The expected output is:

\ding{45}~[EXPECT]
\end{mybox}

\ding{184}~For a bug with a Wrong Answer (WA) error, apart from the input and expected output, the prompt also contains the actual output of the buggy program.
The prompt is described as follows, where [OUTPUT] represents the output of the buggy program given the input filled in "[INPUT]".

\begin{mybox}{Wrong Answer Error Prompt}
There's a bug in the program below. Try to fix it and return the complete fix for the code in the form of the markdown code block.

\ding{45}~[CODE]

The following input triggers a Wrong Answer error:

\ding{45}~[INPUT]

The expected output is:

\ding{45}~[EXPECT]

The actual output is:

\ding{45}~[OUTPUT]
\end{mybox}

\textbf{$\bullet$~Bug localization} show the suspicious lines where bugs are localized in~\cite{wong2016survey}. 
We mark the buggy lines with a comment \texttt{"//bug"} and then ask ChatGPT to repair the code with a bug position. 
\revise{To obtain the buggy line, we compare the lines of every pair of bug and fix. In the bug, the first line that is different from the line in the fix is considered as the buggy line.}
We add another short prompt just following the basic prompt to tell ChatGPT what the comment means.
The prompt is designed as follows, where [CODE] represents the buggy code whose buggy line is followed by the comment \texttt{"//bug"}.

\begin{mybox}{Bug Localization Prompt}
There's a bug in the program below. Try to fix it and return the complete fix for the code in the form of the markdown code block. The location of the bug is in or near the line with a comment "\texttt{//bug}".

\ding{45}~[CODE]
\end{mybox}

\begin{figure}[htbp]
    \centering
    \includegraphics[width=0.9\linewidth]{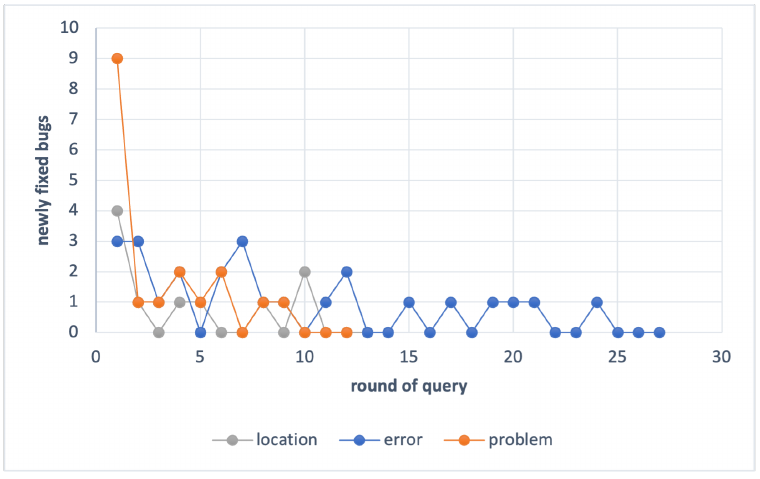}
    \caption{Number of bugs newly fixed in each round with three types of prompts}
    \label{fig:rq2_compare}
\end{figure}

\textbf{\emph{Results.}}
In the 42 bugs that are not fixed in RQ1, 25, 18, and 10 more bugs are fixed when we separately add error information, problem description, and bug location to the basic prompt. 
This shows the promoting effect that more concrete prompts have on the repair performance of ChatGPT. 
Like in RQ1, we query ChatGPT several times until no bugs are newly fixed in continuously three rounds. 
Figure~\ref{fig:rq2_compare} shows how many additional bugs are fixed in every round. 
It only takes 12 and 13 rounds to prompt with problems and bug locations, but 27 rounds are executed when it comes to error information, notably more than the other two prompts. 
The possible reason is that the former two prompts help ChatGPT to focus on a smaller range of code, so the patches it generates in every round are relevantly stabler. 
By contrast, there is vacillation in the patches when error messages are provided, since an error could be caused by different parts of the code. 

\begin{figure}[htbp]
    \centering
    \includegraphics[width=0.8\linewidth]{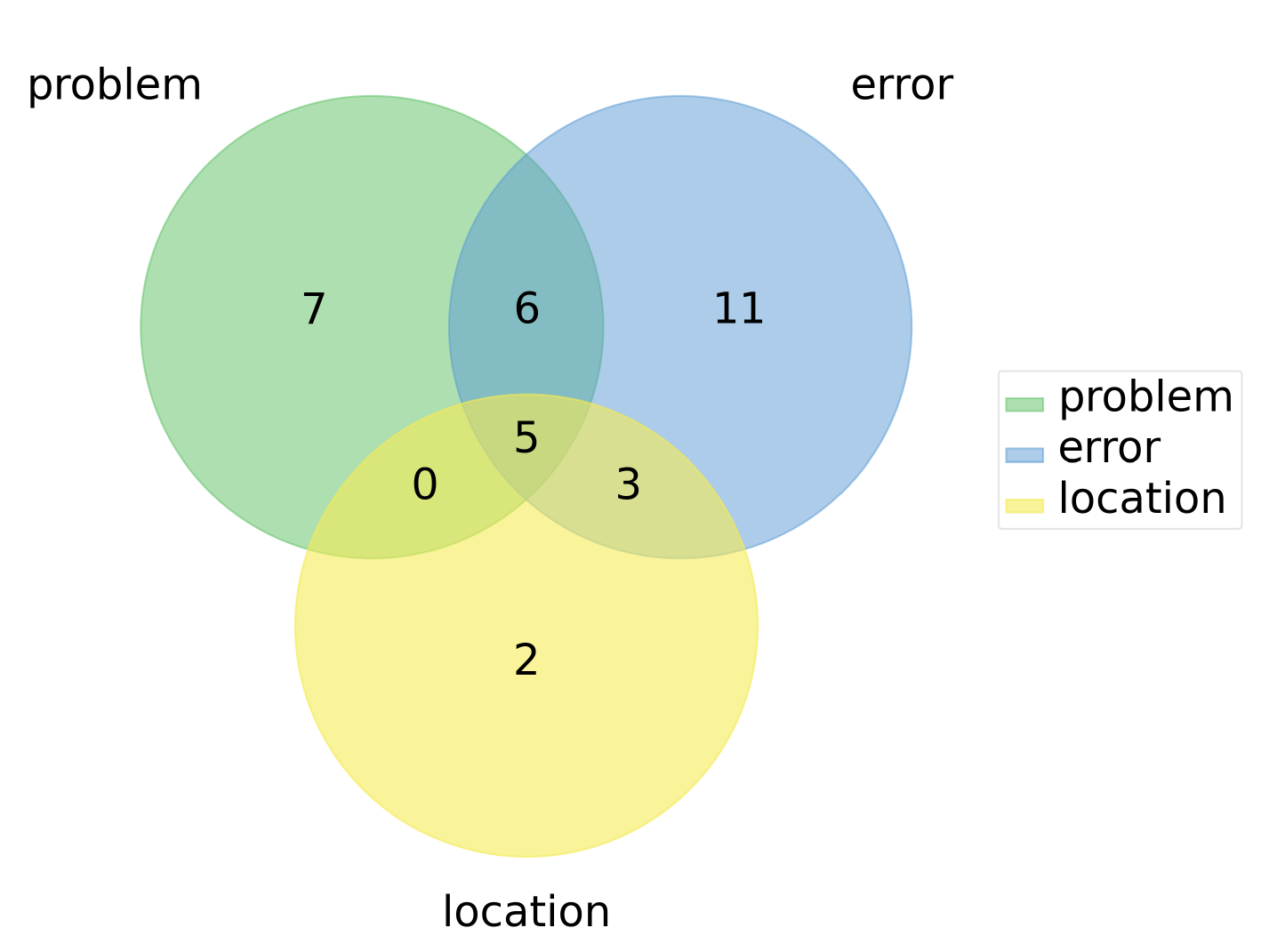}
    \caption{The overlaps of the bugs fixed by three types of prompts}
    \label{fig:rq2_overlap}
\end{figure}

Figure~\ref{fig:rq2_overlap} presents the overlapping relationship among the bugs fixed with the three types of prompts. 
Altogether, 34 bugs are fixed, while only 5 of them are fixed by all three prompts.
Each prompt fixes 11, 7, and 2 bugs that the other two prompts fail to fix, indicating that different prompts contribute disparately to a successful bug fix. 

\textbf{Case Study with Problem Description.}
ChatGPT gains knowledge about the purpose of the program through problem descriptions, so it can find out which part of the code is inconsistent with what the coder is actually trying to do. 
The following is a programming problem and a buggy submission which is fixed by ChatGPT.

\begin{center}
\footnotesize
\fcolorbox{gray!10}{gray!10}{\parbox{.95\linewidth}{

\textbf{Problem Statement:}

Takahashi turned on a computer at time $0$ and clicked the mouse N times. The $i$-th ($1\leq i\leq N$) click was at time $T_i$.

If he consecutively clicked the mouse at time $x_1$ and time $x_2$ (where $x_1$<$x_2$ ), a double click is said to be fired at time $x_2$ if and only if $x_2 - x_1 \leq D$.

What time was a double click fired for the first time? If no double click was fired, print $-1$ instead.

\textbf{Input:}

The input is given from Standard Input in the following format:

$N~D$

$T_1$ $T_2$ ... $T_N$

\textbf{Output:}

If at least one double click was fired, print the time of the first such event; otherwise, print $-1$.

}}
\end{center}

\begin{lstlisting}[caption={An example of fixed bug with problem description},label={listing:pd},float]
    ...
    for(int i = 1; i < count; i++) {
    int x1 = numlist.get(i - 1);
    int x2 = numlist.get(i);
    int dis = x2 - x1;
-   if(dis < distance) {
+   if(dis <= distance) {
        System.out.println(x2);
        break;
    }
    counter++;
    }
    ...
\end{lstlisting}

In Listing~\ref{listing:pd}, in the if statement, the operator that the problem requires is a \texttt{"<="}, \revise{as it is stated in the problem that "a double click is said to be fired at time $x_2$ if and only if \bm{$x_2 - x_1 \leq D$}",}
but the programmer wrongly uses a \texttt{"<"}. 
ChatGPT realizes the discordance between the problem and the function code, and replaces the incorrect operator to have the bug fixed.

\textbf{Case Study with Error Information.}
ChatGPT can also infer where the bug is according to information on the type of the bug as well as on what input the program fails.
Listing~\ref{listing:ei} presents a bug that can only be fixed by {\gpt} with error information. 
To fix the following bug, ChatGPT is informed that for the input "9 737738327422964222", the expected output is "81970925269218254" but the program actually outputs "1251275726". From the prompt, ChatGPT can probably infer that there is an overflow of the output variable "cnt", so it changes the variable type from int to long.

\begin{lstlisting}[caption={An example of fixed bug with error information},label={listing:ei},float]
    ...
    Scanner sc = new Scanner(System.in);
    long A = sc.nextLong();
    long B = sc.nextLong();
-           int cnt = 0;
+           long cnt = 0;
    while(A!=B){
        if(A>B){
            long div = A/B;
            A = A-B*div;
            if(A==0){
                div += -1;
                cnt += div;
                break;
    ...
\end{lstlisting}

\textbf{Case Study with Bug Localization.}
With fault locations, ChatGPT fixes relatively fewer bugs than with the other two types of prompts. 
This could be attributed to the less plentiful information provided by simply buggy lines. Nevertheless, the bug position helps in some circumstances. 
Listing~\ref{listing:bl} presents a bug that is correctly fixed only when the bug location is offered in the prompt. 
While attempting to fix it based on the programming problem or the error message, ChatGPT makes the same mistake: it not only swaps the method \texttt{"nextInt()"} to \texttt{"nextLong()"} but also modifies the output \texttt{"sum - 1"} to \texttt{"sum"} at the end of the program. 
The prompt with the location of the bug in some ways narrows the scope of code that ChatGPT tries to edit, thus deterring ChatGPT from changing the initial correct line.

\begin{lstlisting}[caption={An example of fixed bug with bug localization},label={listing:bl},float]
    ...
    Scanner sc=new Scanner(System.in);
-   long A=sc.nextInt();
+   long A=sc.nextLong();
-   long B=sc.nextInt();
+   long B=sc.nextLong();
    long sum=0;
    while(A!=0 &&B!=0) {
        if (A<B) {
            long tmp=A;
            A=B;
            B=tmp;
        }
        sum+=A/B;
        A=A%B;
    ...
\end{lstlisting}

\myfinding{2}{
The performance under different prompts demonstrates that,
{\gpt} can benefit from more advanced prompts with additional information.
For example, compared with the basic prompt, 25, 18, and 10 more bugs can be fixed with error information, problem description, and buggy lines.
}

\subsection{RQ3: Dialogue Study}
\label{dialogue}

\textbf{\emph{Design.}}
During the conversation with ChatGPT, it is aware of the previous dialogues, and the response depends on both the current prompt and the context of the conversation. 
According to existing work~\cite{cao2023study}, ChatGPT can repair more program faults through performing more dialogues. 
We raise this RQ to investigate the effect of dialogues on ChatGPT's repairing performance.

We focus on the unfixed bugs when giving ChatGPT either the basic prompt or the advanced prompt with error information. 
First, we prompt ChatGPT with the original bug and what type of error it has, and get a patch, which is then tested on the test cases. 
If it fails to pass any of the test cases again, we continue to conduct the next round of dialogue, in which we tell {\gpt} the failure triggered by the patch.
In particular, with different types of bugs, the dialogue prompts are also different. 
Similar to the error information prompts described in Section~\ref{rq2}, when a TLE or RE bug occurs, the [OUTPUT] will not be provided; 
when the bug is a CE, the dialogue will not involve any of the three elements (i.e. [INPUT], [EXPECT] and [OUTPUT]), while with a WA, all of them will be added to the prompt, shown as follows.

\begin{mybox}{Dialogue Prompt}
There's still a Compilation Error/ Time Limit Exceeded Error/ Runtime Error/ Wrong Answer Error in your code triggered by the input:

\ding{45}~[INPUT]

The expected output is:

\ding{45}~[EXPECT]

The actual output is:

\ding{45}~[OUTPUT]

Try to fix it again and return the complete fix for the code.    
\end{mybox}

\revise{

}

The dialogue is continued until ChatGPT successfully repairs the bug or the round of dialogues reaches five. The API of ChatGPT has a limited length of input, so sometimes the dialogue may become too long to process. 
In this case, we simply delete the second round of dialogue since the first round contains the basic and essential bug information. The above process is repeated until no more new bugs are fixed for successive three times.

\textbf{\emph{Results.}}
Through dialogues, ChatGPT fixes 9 bugs among the 17 bugs that keep unfixed when only a single prompt is provided, proving that dialogues can help ChatGPT positively in repairing bugs. This can be because that although ChatGPT mistakenly fixes the bug at the first time, the knowledge towards the right fix is accumulated when we respond to a patch with what problems it has. In this way, ChatGPT can learn from its previous errors and be directed to the true location of the bug, thus the probability of generating a correct patch is improved.

There are mainly three scenarios in which ChatGPT can rectify the wrong patch with dialogues: 
(1) ChatGPT wrongly identifies the position of the bug, so the patch faces the same problem as what the buggy code has. After notifying ChatGPT of the problem that still exists, a new line can be located, which might be where the bug is really in. 
(2) ChatGPT fixes the initial problem, but there is another bug in the code that ChatGPT fails to notice. This usually happens when several bugs lie in different locations of the code. For example, in the first round of dialogue, we give a bug with a Compilation Error, caused by a syntax problem. ChatGPT corrects the syntax but a Wrong Answer problem arises. In this case, ChatGPT is told about the wrong answer and manages to find the other bug. 
(3) ChatGPT finds multiple suspicious code snippets, only part of which are bugs. While fixing the bugs, ChatGPT also turns some originally correct code into the wrong one. From further dialogues, ChatGPT can realize its previous faults and produce the correct patch.

\myfinding{3}{
The performance under a dialogue study demonstrates that,
{\gpt} can repair more difficult-to-fix bugs with dynamic execution feedback in an interaction manner, \eg 9 bugs that have not been fixed in previous prompts are fixed successfully.
}

\section{Discussion}
In the above work, we have demonstrated that ChatGPT has an outstanding performance in program repair. However, we only ask ChatGPT to repair bugs written by human programmers in a coding contest. 
In this section, we aim to investigate on whether ChatGPT can fix bugs in the code generated by itself.
If ChatGPT is able to fix bugs in its own code through dialogues, it will take less effort to debug manually, and help to improve humans' coding efficiency while working with ChatGPT.

To research on this question, we use the method similar to the one in RQ3, but this time the programs to be fixed by ChatGPT are given by itself. 
We first query ChatGPT with the problem descriptions in two AtCoder contests (\ie abc297 and abc298) with the following prompt:

\begin{mybox}{Code Generation Prompt}
\ding{46}~Use Java to solve the following problem. The class name must be \texttt{'Main'}. return the code in the form of markdown code block.

\ding{45}~[PROBLEM]
\end{mybox}

ChatGPT is asked with this prompt for 10 times and generates 10 pieces of code for each coding problem. We run them on all the test cases, and pick out the one that passes the most cases. 
If it passes all the test cases, it will not go into the next step. Otherwise, it is considered as the buggy code to be fixed.
Next, we ask ChatGPT to fix the bug it has previously generated using the same way as in the dialogue study mentioned in Section~\ref{dialogue}. The dialogue round is set as 30 as we find that ChatGPT is not able to fix any bugs within a small number of dialogues, such as 5 or 10.

Among the 16 problems in the two contests, ChatGPT generates correct solutions for 3 of them when first asked to solve the problems, and then it is required to fix the remaining 13 bugs. Despite of ChatGPT's impressive bug-fixing performance in the previous experiments, it is unexpected that only 2 bugs are fixed even though we have performed 30 rounds of dialogues. This indicates that ChatGPT may have limited ability in self-repair.
After careful analysis, the possible reason lies in that the edit actions to fix the developer-submitted programs are minimal (\eg less than 6 token differences in Section~\ref{sec:dataset}), while the initial programs generated by {\gpt} are far from the correct programs and require more edits.

\section{Challenge and Opportunity}

\textbf{Better Prompt Engineering.}
In the experiment, we design different prompts to feed {\gpt} with the buggy code and detailed debugging information.
The results show that {\gpt} is able to fix an impressive number of bugs with the basic prompt in RQ1.
We also find advanced prompts in RQ2 and RQ3 can further boost repair performance.
The quality of a prompt depends on how much information it contains and how well-crafted it is. 
In the bug-fixing scenario, a prompt well-crafted should contain the following element: the instruction to describe the specific task the model needs to perform, context to describe additional information that steers LLMs to better responses, and examples to describe the concrete type of the input and output samples.
In the future, it is important to explore better prompts that effectively guide such LLMs to generate accurate and helpful patches for the buggy code.

\textbf{Domain-specific LLMs.}
The investigated {\gpt} has shown outstanding performance in repairing software bugs, outperforming code domain-specific LLMs, \eg CodeT5.
We think the benefits of {\gpt} lies in a wide range of training data from the internet, including books, articles, websites, and other textual data.
Such diverse training data gives {\gpt} a broad understanding of language and general knowledge, while it also includes a substantial amount of information unrelated to source code and requires a huge parameter size.
On the other hand, domain-specific LLMs are typically trained specifically on datasets relevant to code within a particular domain.
Such domain-specific models are more focused and tailored to the code-related task and can capture domain-specific code transformation patterns, potentially resulting in more accurate patches.
Overall, general-purpose LLMs (\eg~{\gpt}) benefit from their broad exposure to diverse language data, enabling them to offer general programming assistance, while domain-specific LLMs (\eg CodeT5), trained solely on code-related datasets, can provide more specialized and domain-specific guidance. 
In the future, it is interesting to explore the advantages and disadvantages of these two types of LLMs. 
Besides, future work can advance the repair capabilities of LLMs by combining general language understanding with specialized domain knowledge.

\textbf{ChatGPT IDE Integration.}
In our work, we evaluate the performance of {\gpt} with a well-constructed benchmark in terms of the number of fixed software bugs.
\revise{Although {\gpt} is able to fix a considerable number of bugs, it is unclear how such LLMs perform in assisting developers in real-world development environments.}
Different from the existing APR pipeline that directly provides patched code, integrating LLMs like ChatGPT into IDEs can offer benefits.
For example, based on the natural language and programming language understanding capabilities of {\gpt}, developers can describe the behaviors of a bug in natural language, and {\gpt} is able to analyze the buggy program and provide feasible solutions under an iterative interactive process.
Besides, {\gpt} can accept more information about the bug, and even additional details or steps to reproduce the bug in IDE, enabling a more accurate diagnosis.
In contrast, with limited code understanding capabilities, most existing APR techniques focus on accepting buggy code as input, leading to numerous ineffective candidate patches (\eg 1000 candidate patches per bug in CIRCLE~\cite{yuan2022circle}).
In the future, more works are recommended to explore how LLMs streamline the typical bug-fixing workflow and offer valuable insights.

\section{Threats to Validity}
The first threat to validity lies in the repair benchmarks.
We construct \benchmark{} from AtCoder, a programming contest platform.
The collected programs are small-size algorithms, which may not accurately reflect real-world professional software repair capabilities.
There is an increasing trend in using competitive programming as benchmarks in APR, as well as other tasks, such as the popular CodeNET benchmark. 
Besides, {\benchmark} contains programs with varying difficulty and types, written by programmers from diverse backgrounds. Therefore, {\benchmark} can effectively evaluate the repair capabilities of LLMs and foster future work on APR.

The second threat to validity comes from the compared approaches.
In RQ1, we select CodeT5 and PLBART as the baselines to evaluate the effectiveness of {\gpt}. 
We do not consider (1) other LLMs (\eg CodeBERT~\cite{feng2020codebert} and GraphCodeBERT~\cite{guo2020graphcodebert}) because the selected two LLMs represent stat-of-the-art in bug-fixing~\cite{lu2021codexglue};  and (2) existing APR techniques (\eg CIRCLE~\cite{yuan2022circle} and CoCoNut~\cite{lutellier2020coconut}) because our work focuses on LLMs on SE. 
However, considering that our work mainly focuses on empirical
evaluations, the improvement of {\gpt} over the baselines is enough to demonstrate the promising future of boosting program repair on top of {\gpt}.

The third threat to validity is the selection of {\gpt} and APR.
In our work, we regard data leakage as a common issue that may appear in a variety of SE tasks involving black-box LLMs.
We only conduct experiments to evaluate the capabilities of {\gpt} in repairing software bugs,
Thus, our findings may not be generalizable to other LLMs and tasks.
Considering the fact that (1) ChatGPT is one of the state-of-the-art LLMs and has been extensively studied in recent works, and (2) APR plays a vital role in software development, and a number of APR works leverage LLMs to generate patches, We believe that ChatGPT and APR can indeed serve as representative examples of LLMs and SE tasks, respectively.

\section{Related Work}
\label{sec:rw}

\subsection{Automated Program Repair}
Existing APR techniques are generally divided into traditional and learning-based ones~\cite{bader2019getafix,jiang2018shaping,ghanbari2019practical}. 
Traditional APR, especially template-based APR, is proven to perform well in fixing bugs.
For example, PAR~\cite{kim2013automatic} first proposes a patch generation method based on fix patterns, which are drawn from over 60,000 patches written manually.
TBar~\cite{liu2019tbar} systematically summarizes frequently-used fix templates from the literature, and applies them in fixing program bugs.
There are different ways to obtain fix templates.
For example, In AVATAR \cite{Liu2019Avatar}, fix patterns come from code changes that can address violations in static bug detection tools.
FixMiner \cite{koyuncu2020fixminer} develops an automated template mining tool, in which a clustering strategy is applied to mine code changes from bugs and patches.

Recently, learning-based APR has been proposed to transform buggy code into the correct one automatically.
For example, DLFix \cite{li2020dlfix} raises a two-layer deep learning model to learn code transformation from bug fixes and surrounding contexts.
CURE \cite{jiang2021cure} pre-trains a programming language model based on a large codebase, and optimizes the searching efficiency by proposing a novel search strategy as well as using a subword tokenization technique. 
In this paper, we do not include these techniques as baselines because (1) previous studies~\cite{xia2022less,jiang2023impact} demonstrate existing LLMs can outperform  APR approaches; and (2) as an empirical study, our work mainly focuses on LLMs and select two state-of-the-art LLMs.

\subsection{Large Language Models in SE}
There is growing interest in leveraging LLMs in SE tasks~\cite{niu2022spt,tufano2019learning}. 
For example, Zeng et al. \cite{zeng2022extensive} perform an extensive study that evaluates eight pre-trained LLMs (\eg CodeBERT \cite{feng2020codebert}, CodeT5 \cite{wang2021codet5}, and GraphCodeBERT \cite{guo2020graphcodebert}) on seven program understanding and generation tasks, and compare the pre-trained models with non-pre-trained domain-specific techniques, demonstrating that pre-trained models significantly perform better in code understanding than other state-of-the-art techniques. 
In the field of program repair, LLMs are exploited to directly generate patches so as to break the limitations in learning-based techniques~\cite{yuan2022circle,xia2022less,xia2022practical}.
AlphaRepair~\cite{xia2022less} introduces the mask prediction task of the pre-trained model CodeBERT \cite{feng2020codebert} into APR, and implements a cloze-style APR tool without using any bug-fixing historic code.
Prenner et al.~\cite{prenner2022can} research on the program repair ability of Codex, a GPT-3-based language model aiming to translate natural language to programming language. 
Although Codex is not specifically trained for APR tasks, it is still effective at fixing bugs, especially those in Python language on the benchmark QuixBugs \cite{lin2017quixbugs}.

Recently, ChatGPT is attracting a lot of attention because of its stunning ability of understanding and responding to conversations started by humans. In APR, ChatGPT has been proven to have outstanding performance in fixing bugs from popular datasets (e.g. Defects4J \cite{just2014defects4j} and QuixBugs~\cite{lin2017quixbugs}). 
Sobania et al. \cite{sobania2023analysis} analyze ChatGPT's program repair behavior through both giving a single request and conducting more discussions with ChatGPT. Cao et al. \cite{cao2023study} focus on studying ChatGPT's capability in fixing deep learning programs. 
Xia et al. \cite{xia2023keep} propose an APR approach based on ChatGPT that fully utilizes conversations by offering instant feedbacks about previous patches. 
In this work, we review the data leakage issue of black-box LLMs by taking {\gpt} and APR as examples.

\section{Conclusion}
\label{sec:con}
In this paper, we seek to review the overlooked data leakage issue of black-box LLMs in the SE domain.
In particular, we evaluate {\gpt}'s capability of program repair on a clean dataset {\benchmark}.
The results demonstrate that ChatGPT generates 109 correct patches over 151 bugs when only given the basic prompt.
Besides, {\gpt} continues to fix 18, 25, and 10 additional bugs with prompts containing programming problem descriptions, error messages, and bug localizations. 
Through engaging in dialogues, nine more bugs are fixed by {\gpt}.
These results indicate that ChatGPT has a promising bug-fixing ability, which can be further enhanced by proper prompts and more dialogues.

\bibliographystyle{ACM-Reference-Format}
\bibliography{sample-base}

\end{document}